\documentclass{article}
\usepackage{etex}
\usepackage[margin=1.5in]{geometry}
\usepackage{mathtools}
\usepackage{xcolor}
\usepackage{enumitem,amsmath,amssymb}
\usepackage{url}
\usepackage[linesnumbered,boxed,noline,noend]{algorithm2e}

\usepackage{tikz}
\usetikzlibrary{arrows}

\usepackage{subfig}
\usepackage{array,booktabs,multirow}
\usepackage{placeins}

\usepackage{logictools}
\usepackage{prooftheory}
\usepackage{comment}
\usepackage{mathenvironments}
\usepackage{drawproof}
\usepackage{bussproofs}
\usepackage{tensor}
\usepackage{mathtools}
\usepackage{amsmath}
\usepackage{amsthm}

\usepackage{graphicx}
\newcommand{\freevar}[1]{\mathrm{Var}(#1)}

\newcommand{\axiom}[1]{\widehat{#1}}
\newcommand{\n}{v}
\newcommand{\raiz}[1]{\rho(#1)}

\DeclarePairedDelimiter{\abs}{\lvert}{\rvert}

\newcommand{\con}[3]{\lfloor #1 \rfloor_{#2}^{#3}}

\newcommand{\res}[4]{\mathrel{\operatorname*{\odot}_{#1 #3}^{#2 #4}}}

\newtheorem{thm}{Theorem}[section]

\newtheorem{example}[thm]{Example}

\theoremstyle{definition}
\newtheorem{definition}{Definition}[section]

\usepackage{authblk}

\begin{document}
\title{Partial Regularization of First-Order Resolution Proofs}

\author[1]{Jan Gorzny}
\author[2]{Ezequiel Postan}
\author[3]{Bruno Woltzenlogel Paleo}
\affil[1]{School of Computer Science, University of Waterloo, 200 University Ave. W., Waterloo, ON N2L 3G1, Canada}
\affil[2]{Universidad Nacional de Rosario, Av. Pellegrini 250, S2000BTP Rosario, Santa Fe, Argentina}
\affil[3]{Vienna University of Technology, Karlsplatz 13, 1040, Vienna, Austria}

  \maketitle




\begin{abstract}
Resolution and superposition are common techniques which have seen widespread use with propositional and first-order logic in modern theorem provers. In these cases, resolution proof production is a key feature of such tools; however, the proofs that they produce are not necessarily as concise as possible.
For propositional resolution proofs, there are a wide variety of proof compression techniques. There are fewer techniques for compressing first-order resolution proofs generated by automated theorem provers.
This paper describes an approach to compressing first-order logic proofs based on lifting proof compression ideas used in propositional logic to first-order logic. One method for propositional proof compression is \emph{partial regularization}, which removes an inference $\eta$ when it is redundant in the sense that its pivot literal already occurs as the pivot of another inference in every path from $\eta$ to the root of the proof. 
This paper describes the generalization of the partial-regularization algorithm
\RecyclePivotsIntersection \cite{LURPI}
from propositional logic to first-order logic. The generalized algorithm performs partial regularization of resolution proofs containing resolution and factoring inferences with \emph{unification}. 
An empirical evaluation of the generalized algorithm and its combinations with the previously lifted \SFOLowerUnits algorithm
\cite{GFOLU} is also presented.

\end{abstract}





\sloppy  

\setcounter{footnote}{0}
\newcommand{\la}{\leftarrow}

\section{Introduction} 

First-order automated theorem provers, commonly based on refinements and extensions of resolution and superposition calculi \cite{Vampire,EProver,Spass,spassT,Beagle,cruanes2015extending,prover9-mace4}, have recently achieved a high degree of maturity. Proof production is a key feature that has been gaining importance, as proofs are crucial for applications that require certification of a prover's answers or that extract additional information from proofs (e.g. unsat cores, interpolants, instances of quantified variables). Nevertheless, proof production is non-trivial \cite{SchulzAPPA}, and the most efficient provers do not necessarily generate the shortest proofs. One reason for this is that efficient resolution provers use refinements that restrict the application of inference rules. Although fewer clauses are generated and the search space is reduced, refinements may exclude short proofs whose inferences do not satisfy the restriction.

Longer and larger proofs take longer to check, may consume more memory during proof-checking and occupy more storage space, and may have a larger unsat core, if more input clauses are used in the proof, and a larger Herbrand sequent, if more variables are instantiated \cite{B10,B12,B16,ResolutionHerbrand,Reis}. For these technical reasons, it is worth pursuing efficient algorithms that compress proofs after they have been found. Furthermore, the problem of proof compression is closely related to Hilbert's 24th Problem \cite{Hilbert24Problem}, which asks for criteria to judge the simplicity of proofs. Proof length is arguably one possible criterion for some applications.

For propositional resolution proofs, as those typically generated by SAT- and SMT-solvers, there is a wide variety of proof compression techniques. Algebraic properties of the resolution operation that are potentially useful for compression were investigated in \cite{bwp10}.
Compression algorithms based on rearranging and sharing chains of resolution inferences have been
developed in \cite{Amjad07} and \cite{Sinz}.  Cotton \cite{CottonSplit} proposed an algorithm that
compresses a refutation by repeatedly splitting it into a proof of a heuristically chosen literal $\ell$
and a proof of $\dual{\ell}$, and then resolving them to form a new refutation.  The {\ReduceReconstruct} algorithm \cite{RedRec} searches for locally redundant
subproofs that can be rewritten into subproofs of stronger clauses and with fewer resolution steps.
Bar-Ilan et al. \cite{RP08} and Fontaine et al. \cite{LURPI} described a linear time proof compression algorithm based on partial
regularization, which removes an inference $\eta$ when it is redundant in the sense that its pivot literal already occurs as the pivot of another inference in every path from $\eta$ to the root of the proof.

In contrast, although proof output has been a concern in first-order automated reasoning for a longer time than in propositional SAT-solving, there has been much less work on simplifying first-order proofs. For tree-like sequent calculus proofs, algorithms based on cut-introduction \cite{BrunoLPAR,Hetzl} have been proposed. However, converting a DAG-like resolution or superposition proof, as usually generated by current provers, into a tree-like sequent calculus proof may increase the size of the proof. For arbitrary proofs in the Thousands of Problems for Theorem Provers (TPTP) \cite{TPTP} format (including DAG-like first-order resolution proofs), there is an algorithm \cite{LPARCzech} that looks for terms that occur often in any Thousands of Solutions from Theorem Provers (TSTP) \cite{TPTP} proof and abbreviates them.

The work reported in this paper is part of a new trend that aims at lifting successful propositional proof compression algorithms to first-order logic. Our first target was the propositional {\LowerUnits} ({\LU}) algorithm \cite{LURPI}, which delays resolution steps with unit clauses, and we lifted it to a new algorithm that we called
{\SFOLowerUnits} 
({\GFOLU}) algorithm \cite{GFOLU}. Here we continue this line of research by lifting the 
\texttt{Recycle\-PivotsWithIntersection}
({\RPI}) algorithm \cite{LURPI}, which improves the \texttt{RecyclePivots} ({\RP}) algorithm \cite{RP08} by detecting nodes that can be regularized even when they have multiple children. 

Section \ref{sec:res} introduces the well-known first-order resolution calculus with notations that are suitable for describing and manipulating proofs as first-class objects. 
Section \ref{Section:RPI} summarizes the propositional {\RPI} algorithm.
Section \ref{sec:Challenges} discusses the challenges that arise in the first-order case (mainly due to unification), which are not present in the propositional case, and conclude with conditions useful for first-order regularization. Section \ref{sec:FORPI} describes an algorithm that overcomes these challenges. Section \ref{sec:exp} presents experimental results obtained by applying this algorithm, and its combinations with {\GFOLU}, on hundreds of proofs generated with the {\SPASS} theorem prover on TPTP benchmarks \cite{TPTP} and on randomly generated proofs. Section \ref{sec:conclusion} concludes the paper.

It is important to emphasize that this paper targets proofs in a pure first-order resolution calculus (with resolution and factoring rules only), without refinements or extensions, and without equality rules. As most state-of-the-art resolution-based provers use variations and extensions of this pure calculus and there exists no common proof format, the presented algorithm cannot be directly applied to the proofs generated by most provers, and even {\SPASS} had to be specially configured to disable {\SPASS}'s extensions in order to generate pure resolution proofs for our experiments. By targeting the pure first-order resolution calculus, we address the common theoretical basis for the calculi of various provers. In the Conclusion (Section \ref{sec:conclusion}), we briefly discuss what could be done to tackle common variations and extensions, such as splitting and equality reasoning. Nevertheless, they remain topics for future research beyond the scope of this paper.

\section{The Resolution Calculus}
\label{sec:res}

As usual, our language has infinitely many variable symbols (e.g. $x$, $y$, $z$, $x_1$, $x_2$, \ldots), constant symbols (e.g. $a$, $b$, $c$, $a_1$, $a_2$, \ldots), function symbols of every arity (e.g $f$, $g$, $f_1$, $f_2$, \ldots) and predicate symbols of every arity (e.g. $P$, $Q$, $P_1$, $P_2$,\ldots). A \emph{term} is any variable, constant or the application of an $n$-ary function symbol to $n$ terms.
An \emph{atomic formula} (\emph{atom}) is the application of an $n$-ary predicate symbol to $n$ terms. A \emph{literal} is an atom or the negation of an atom. The
\emph{complement} of a literal $\ell$ is denoted $\dual{\ell}$ (i.e. for any atom $P$,
$\dual{P} = \neg P$ and $\dual{\neg P} = P$). The \emph{underlying atom} of a literal $\ell$ is denoted $\abs{\ell}$ (i.e. for any atom $p$, $\abs{P} = P$ and $\abs{\neg P} = P$). A
\emph{clause} is a multiset of literals. $\bot$ denotes the \emph{empty clause}. A \emph{unit clause} is a clause with a single literal. Sequent notation is used for clauses (i.e. $P_1,\ldots,P_n \seq Q_1,\ldots, Q_m$ denotes the clause $\{ \neg P_1,\ldots, \neg P_n, Q_1, \ldots, Q_m \}$).
$\freevar{t}$ (resp. $\freevar{\ell}$, $\freevar{\clause}$) denotes the set of variables in the term $t$ (resp. in the literal $\ell$ and in the clause $\clause$).
A \emph{substitution} $\{ x_1\backslash t_1, x_2 \backslash t_2, \ldots \}$ is a mapping from variables $\{ x_1, x_2, \ldots \}$ to, respectively, terms $\{t_1, t_2, \ldots \}$. The application of a substitution $\sigma$ to a term $t$, a literal $\ell$ or a clause $\clause$ results in, respectively, the term $t \sigma$, the literal $\ell \sigma$ or the clause $\clause \sigma$, obtained from $t$, $\ell$ and $\clause$ by replacing all occurrences of the variables in $\sigma$ by the corresponding terms in $\sigma$. A literal $\ell$ \emph{matches} another literal $\ell'$ if there is a substitution $\sigma$ such that $\ell\sigma=\ell'$. A \emph{unifier} of a set of literals is a substitution that makes all literals in the set equal. We will use $X \sqsubseteq Y$ to denote that $X$ \emph{subsumes} $Y$, when there exists a substitution $\sigma$ such that $X\sigma \subseteq Y$.

The resolution calculus used in this paper has the following inference rules: 


\begin{definition}[Resolution] \label{def:fores} \hfill
\begin{prooftree}
\AxiomC{$\eta_1$: $\Gamma_L' \cup \{\ell_L\}$ }
\AxiomC{$\eta_2$: $\Gamma_R'\cup \{\ell_R\}$ }
\BinaryInfC{$\psi$: $\Gamma_L'\sigma_L \cup \Gamma_R'\sigma_R$}
\end{prooftree}
where $\sigma_L$ and $\sigma_R$ are substitutions such that $\ell_L\sigma_L=\dual{\ell_R}\sigma_R$. The literals $\ell_L$ and $\ell_R$ are \emph{resolved literals}, whereas $\ell_L \sigma_L$ and $\ell_R \sigma_R$ are its \emph{instantiated resolved literals}. The \emph{pivot} is the underlying atom of its instantiated resolved literals (i.e. $\abs{\ell_L \sigma_L}$ or, equivalently, $\abs{\ell_R \sigma_R}$).
\end{definition}

\begin{definition}[Factoring] \label{def:fofact} \hfill
\begin{prooftree}
\AxiomC{$\eta_1$: $\Gamma' \cup \{\ell_1,\ldots,\ell_n\}$ }
\UnaryInfC{$\psi$: $\Gamma'\sigma \cup \{\ell\}$}
\end{prooftree}
where $\sigma$ is a unifier of $\{\ell_1,\ldots,\ell_n\}$ and $\ell=\ell_i\sigma$ for any $i\in \{1,\ldots,n\}$.
\end{definition}

A \emph{resolution proof} is a directed acyclic graph of clauses where the edges correspond to the inference rules of resolution and factoring, as explained in detail in Definition \ref{def:proof}. A \emph{resolution refutation} is a resolution proof with root $\bot$.

\begin{definition}[First-Order Resolution Proof] 
\label{def:proof}
A directed acyclic graph $\langle V, E, \clause \rangle$, where $V$ is a set of nodes and $E$ is a
set of edges labeled by literals and substitutions (i.e. $E \subset V \times 2^{\mathcal{L}} \times \mathcal{S} \times V$, where $\mathcal{L}$ is the set of all literals and $\mathcal{S}$ is the set of all substitutions, and $\n_1
\xrightarrow[\sigma]{\ell} \n_2$ denotes an edge from node $\n_1$ to node $\n_2$ labeled by the literal $\ell$ and the substitution $\sigma$), is a
proof of a clause $\clause$ iff it is inductively constructible according to the following cases:
\begin{itemize}
  \item \textbf{Axiom:} If $\Gamma$ is a clause, $\axiom{\Gamma}$ denotes some proof $\langle \{ \n \}, \varnothing,
    \Gamma \rangle$, where $\n$ is a new (axiom) node.
  \item \textbf{Resolution\footnote{This is referred to as ``binary resolution'' elsewhere, with the understanding that ``binary'' refers to the number of resolved literals, rather than the number of premises of the inference rule.}:} If $\psi_L$ is a proof $\langle V_L, E_L, \clause_L \rangle$ and
    $\psi_R$ is a proof $\langle V_R, E_R, \clause_R \rangle$, where $\Gamma_L$ and $\Gamma_R$ satisfy the requirements of Definition \ref{def:fores},
    then
    $\psi_L \res{\ell_L}{\sigma_L}{\ell_R}{\sigma_R} \psi_R$ denotes a proof $\langle V, E, \Gamma \rangle$ s.t.
    \begin{align*}
     \hspace{-0.6cm} V &= V_L \cup V_R \cup \{\n \}    \\
      \hspace{-0.6cm} E &= E_L \cup E_R \cup 
                    \left\{ \raiz{\psi_L} \xrightarrow[\sigma_L]{\{\ell_L\} } \n, 
                            \raiz{\psi_R} \xrightarrow[\sigma_R]{\{\ell_R\} } \n \right\}    \\
    \hspace{-0.6cm}  \Gamma &= \clause_L' \sigma_L \cup  \clause_R' \sigma_R
    \end{align*}
    where $\n$ is a new (resolution) node and $\raiz{\varphi}$ denotes the root node of $\varphi$. 
  \item \textbf{Factoring:}
  If $\psi'$ is a proof $\langle V', E', \clause' \rangle$ such that $\Gamma$ satisfies the requirements of Definition \ref{def:fofact}, then $\con{\psi}{\{\ell_1, \ldots \ell_n\}}{\sigma}$ denotes a proof $\langle V, E, \Gamma \rangle$ s.t.
    \begin{align*}
         \hspace{-0.6cm} V &= V' \cup \{\n \} \\
         \hspace{-0.6cm} E &= E' \cup \{ \raiz{\psi'} \xrightarrow[\sigma]{\{\ell_1, \ldots \ell_n\}} \n \} \\
       \hspace{-0.6cm} \Gamma &= \clause' \sigma \cup \{ \ell \}
    \end{align*}
    where $\n$ is a new (factoring) node, and $\raiz{\varphi}$ denotes the root node of $\varphi$.
  \qed
\end{itemize}
\end{definition}

\begin{example}\label{ex:resolutionproof}

An example first-order resolution proof is shown below.

\begin{scriptsize}
\begin{prooftree}
\def\e{\mbox{\ $\vdash$\ }}
\AxiomC{$\eta_1$: $Q(x),Q(a)\e P(b)$ } 
\AxiomC{$\eta_2$: $P(b)\e$ } 
\BinaryInfC{$\eta_3$: $Q(x),Q(a) \e$}
\UnaryInfC{$\eta_3'$: $Q(a) \e$ \hspace{-1cm}}
\AxiomC{$\eta_2$} 
\AxiomC{\hspace{-0.5cm}$\eta_4$: $\e P(b), Q(y)$ } 
\BinaryInfC{\hspace{-2cm} $\eta_5$: $\e Q(y)$}
\BinaryInfC{$\psi$: $\bot$}
\end{prooftree}
\end{scriptsize}

\noindent
The nodes $\eta_1$, $\eta_2$, and $\eta_4$ are axioms. Node $\eta_3$ is obtained by resolution on $\eta_1$ and $\eta_2$ where $\ell_L = P(b)$, $\ell_R = \neg P(b)$, and $\sigma_L = \sigma_R = \emptyset$. The node $\eta_3'$ is obtained by a factoring on $\eta_3$ with $\sigma=\{x\setminus a\}$. The node $\eta_5$ is the result of resolution on $\eta_2$ and $\eta_4$ with $\ell_L = \neg P(b)$, $\ell_R = P(b)$, $\sigma_L=\sigma_R = \emptyset$. Lastly, the conclusion node $\psi$ is the result of a resolution of $\eta_3'$ and $\eta_5$, where $\ell_L = \neg Q(a)$, $\ell_R = Q(y)$, $\sigma_L = \emptyset$, and $\sigma_R = \{ y \setminus a\}$. The directed acyclic graph representation of the proof (with edge labels omitted) is shown in Figure \ref{fig:dagex}.\\

\begin{figure}[bt]
\begin{centering}
\begin{tikzpicture}
  \tikzstyle{vertex}=[circle,minimum size=10pt,inner sep=0pt]
\tikzset{edge/.style = {->,> = latex'}}

    \node[vertex] (n1) at (-2,1.75) {$\eta_1$};
    \node[vertex] (n2) at (0,1.75) {$\eta_2$};
    \node[vertex] (n4) at (2,1.75) {$\eta_4$};
    \node[vertex] (n5) at (1,0.5) {$\eta_5$};
    \node[vertex] (n3) at (-1,1.25) {$\eta_3$};
    \node[vertex] (n3p) at (-1,0.5) {$\eta_3'$};
    \node[vertex] (r) at (0,0) {$\psi$};

\draw[edge] (r) -- (n5);
\draw[edge] (r) -- (n3p);

\draw[edge] (n3p) -- (n3);

\draw[edge] (n3) -- (n1);
\draw[edge] (n3) -- (n2);

\draw[edge] (n5) -- (n2);
\draw[edge] (n5) -- (n4);
\end{tikzpicture}

\end{centering}
\caption{The proof in Example \ref{ex:resolutionproof}.}
\label{fig:dagex}
\end{figure}
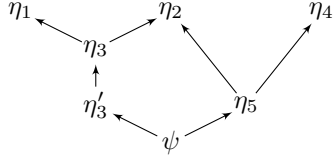
\end{example}

\section{Algorithm {\RecyclePivotsIntersection}}
\label{Section:RPI}

\newcommand{\tRes}{\odot}
\newcommand{\tResFact}{\otimes}
\newcommand{\tResChain}{\ominus}
\newcommand{\AXC}{\AxiomC}
\newcommand{\BIC}{\BinaryInfC}
\newcommand{\RName}[1]{\RightLabel{#1}}
\newcommand{\p}[1]{\hat{#1}}
\newcommand{\ub}[2]{\underbrace{#1}_{#2}}
\newcommand{\tResStar}{\circledast}



This section explains {\RecyclePivotsIntersection} ({\RPI}) \cite{LURPI}, which aims to compress irregular propositional proofs. It can be seen as a simple 
but significant modification of the {\RP} algorithm described in 
\cite{RP08}, 
from which it derives its name. 
Although in the worst case full regularization can increase the proof length exponentially 
\cite{Tseitin}, these algorithms show that 
many irregular proofs can have their length decreased if a careful partial regularization is performed. 

We write $\psi[\eta]$ to denote a \emph{proof-context}
$\psi[\_]$ with a single placeholder replaced by the subproof $\eta$.
We say that a proof of the form $\psi[\eta \tRes_p \psi'[\eta'\tRes_p\eta_2]]$ is \emph{irregular}.

\begin{example}
Consider an irregular proof and assume, without loss of generality, that $p \in \eta$ and $p \in \eta'$, as in the proof of $\psi$ below. The proof of $\psi$ can be written as $(\eta \tRes_p ( \eta_1 \tRes (\eta' \tRes_p \eta'')))$, or $(\eta \tRes_p \psi'[(\eta' \tRes_p \eta'')])$ where $\psi'[(\eta' \tRes_p \eta'')] = (\eta_1 \tRes (\eta' \tRes_p \eta''))$ is the sub-proof of $\lnot p$.
\begin{footnotesize}
\begin{prooftree}

		\AXC{$\eta$: $p$} \RName{$p$}
			\AXC{$\eta_1$: $\lnot r, \lnot p$}
		\AXC{$\eta'$: $p$}
				\AXC{$ \eta''$: $\lnot p, r$} \RName{$p$}
	\BIC{$r$}

	\BIC{$\lnot p$} \RName{$p$}

		\BIC{$\psi$: $\bot$}	
\end{prooftree}
\label{ex:rpi-example-a}
\end{footnotesize}
\noindent
Then, if $\eta' \tRes_p \eta''$ is replaced by $\eta''$ within the proof-context $\psi'[\ ]$, the clause $\eta \tRes_p \psi'[\eta'']$ subsumes the clause $\eta \tRes_p \psi'[\eta' \tRes_p \eta'']$, because even though the literal $\neg p$ of $\eta''$ is
propagated down, it gets resolved against the literal $p$ of $\eta$ later on below in the proof. More precisely, even though it might be the case that $\neg p \in \psi'[\eta'']$ while $\neg p \notin \psi'[\eta' \tRes_p \eta'']$, it is necessarily the case that $\neg p \notin \eta \tRes_p \psi'[\eta' \tRes_p \eta'']$ and $\neg p \notin \eta \tRes_p \psi'[\eta'']$. In this case, the proof can be regularized as follows.

\begin{footnotesize}
\begin{prooftree}

		\AXC{$\eta$: $p$}
			\AXC{$\eta_1$: $\lnot r, \lnot p$}

				\AXC{$ \eta''$: $\lnot p, r$}

	\BIC{$\lnot p$} \RName{$p$}

		\BIC{$\psi$: $\bot$}	

\end{prooftree}
\end{footnotesize}
\end{example}

\begin{figure*}%
\centering
\subfloat[A propositional proof before compression by {\RPI}.]{%
\begin{footnotesize}
\AXC{$ \eta_1 $}
		\AXC{$ \eta_2: a, c, \neg b $}
				\AXC{$ \eta_1: \neg a$}
						\AXC{$ \eta_3:  a, b $} \RName{$a$}
					\BIC{$ \eta_4: b$} \RName{$b$}
			\BIC{$ \eta_5: a, c$}	\RName{$a$}
	\BIC{$\eta_6: c$}
		\AXC{$ \eta_4 $}
				\AXC{$ \eta_7: a, \neg b, \neg c $} \RName{$b$}
			\BIC{$ \eta_8: a, \neg c$}	\RName{$c$}
					\AXC{$ \eta_1 $}  \RName{$a$}
				\BIC{$ \eta_9: \neg c$}	\RName{$c$}
		\BIC{$\psi: \bot$}	
		 \DisplayProof 
\label{ex:rpi-example-a}
\end{footnotesize}}\\%
\subfloat[A propositional proof after compression by {\RPI}.]{%
\begin{small}
\AXC{$ \eta_1: \neg a $}
		\AXC{$ \eta_2: a, c, \neg b $}
						\AXC{$ \eta_3:  a, b $}\RName{$ $}
			\BIC{$ \eta_5: a, c$}	\RName{$ $}
	\BIC{$\eta_6: c$}
		\AXC{$ \eta_3 $}
				\AXC{$ \eta_7: a, \neg c, \neg b $} \RName{$ $}
			\BIC{$ \eta_8: a, \neg c$}	\RName{$ $}
					\AXC{$ \eta_1 $}  \RName{$ $}
				\BIC{$ \eta_9: \neg c$}	\RName{$ $}
		\BIC{$\psi: \bot$}	
		 \DisplayProof 
\label{ex:rpi-example-b}
\end{small}
}%

\caption{A {\RPI} example.}
\label{ex:rpi-example}
\end{figure*}

Although the remarks above suggest that it is safe to replace $\eta' \tRes_p
\eta''$ by $\eta''$ within the proof-context $\psi'[\ ]$, this is not always the
case. If a node in $\psi'[\ ]$ has a child in $\psi[\ ]$, then the literal $\neg
p$ might be propagated down to the root of the proof, and hence, the clause
$\psi[ \eta \tRes_p \psi'[ \eta''] ]$ might not subsume the clause $\psi[ \eta
\tRes_p \psi'[\eta' \tRes_p \eta''] ]$. Therefore, it is only safe to do the
replacement if the literal $\neg p$ gets resolved in all paths from $\eta''$ to the root or if it already occurs in the root clause of the original proof $\psi[ \eta \tRes_p \psi'[\eta' \tRes_p \eta''] ]$.

\IncMargin{0.5em}
\begin{algorithm}[b]
\begin{footnotesize}
\SetKwInOut{Input}{input}\SetKwInOut{Output}{output}
\SetKwData{units}{unitsQueue}
\SetKwData{fixedUnits}{fixedUnitsQueue}

\Input{A proof $\psi$}
\Output{A possibly less-irregular proof $\psi'$}

\BlankLine

$\psi'$ $\la$ $\psi$\;
traverse $\psi'$ bottom-up and \ForEach{node $\eta$ in $\psi'$}{
   \If{$\eta$ is a resolvent node}{
     setSafeLiterals($\eta$) \;
     regularizeIfPossible($\eta$)
   }
  }
$\psi'$ $\la$ fix($\psi'$) \;
\Return {$\psi'$}\;
\caption{\label{algo:RPI} \texttt{\RPI}}
\end{footnotesize}
\end{algorithm}
\DecMargin{0.5em}

These observations lead to the idea of traversing the proof in a bottom-up
manner, storing for every node a set of \emph{safe literals} that get resolved
in all paths below it in the proof (or that already occurred in the root clause
of the original proof). Moreover, if one of the node's resolved literals belongs
to the set of safe literals, then it is possible to regularize the node by
replacing it by one of its parents (cf.\ Algorithm~\ref{algo:RPI}). 

The regularization of a node should replace a node by one of its parents, and more precisely by the parent whose clause contains the resolved literal that is safe. After regularization, all nodes below the regularized node may have to be fixed. However, since the regularization is done with a bottom-up traversal, and only nodes below the regularized node need to be fixed, it is again possible to postpone fixing and do it with only a single traversal afterwards. 
Therefore, instead of replacing the irregular node by one of its parents immediately, 
its other parent is marked as \texttt{deletedNode}, as shown in Algorithm~\ref{algo:Regularize}. Only later during fixing, 
the irregular node is actually replaced by its surviving parent (i.e. the parent that is not marked as \texttt{deletedNode}).

\IncMargin{0.5em}
\begin{algorithm}[p]
\begin{footnotesize}
\SetKwInOut{Input}{input}\SetKwInOut{Output}{output}
\SetKwData{units}{unitsQueue}
\SetKwData{fixedUnits}{fixedUnitsQueue}

\Input{A node $\eta$}
\Output{nothing (but the proof containing $\eta$ may be changed)}

\BlankLine
    \uIf{$\eta${\upshape.rightResolvedLiteral} $\in$ $\mathcal{S}(\eta)$}{
      mark left parent of $\eta$ as \texttt{deletedNode} \;
      mark $\eta$ as regularized
    }
    \ElseIf{\textrm{$\eta${\upshape.leftResolvedLiteral} $\in$  $\mathcal{S}(\eta)$}}{
      mark right parent of $\eta$ as \texttt{deletedNode} \;
      mark $\eta$ as regularized
    }
\caption{\label{algo:Regularize} \texttt{regularizeIfPossible}}
\end{footnotesize}
\end{algorithm}
\DecMargin{0.5em}

\IncMargin{0.5em}
\begin{algorithm}[p]
\begin{footnotesize}
\SetKwInOut{Input}{input}\SetKwInOut{Output}{output}
\SetKwData{units}{unitsQueue}
\SetKwData{fixedUnits}{fixedUnitsQueue}

\Input{A node $\eta$}
\Output{nothing (but the node $\eta$ gets a set of safe literals)}

\BlankLine

    \uIf{$\eta$ is a root node with no children}{
      $\mathcal{S}(\eta)$ $\la$ $\eta$.clause  
    }
    \Else{
      \ForEach{$\eta'$ $\in$ $\eta${\upshape.children}}{
        \uIf{$\eta'$ is marked as regularized}{ 
          safeLiteralsFrom($\eta'$) $\la$  $\mathcal{S}(\eta')$ \;}
        \uElseIf{$\eta$ is left parent of $\eta'$}{ 
        	safeLiteralsFrom($\eta'$) $\la$  $\mathcal{S}(\eta')$ $\cup$ 
        	\hspace{6cm} $~~~~$ \{ $\eta'$.rightResolvedLiteral \} \;
        }
        \ElseIf{$\eta$ is right parent of $\eta'$}{ 
			safeLiteralsFrom($\eta'$) $\la$ $\mathcal{S}(\eta')$ $\cup$ 
			\hspace{6cm} $~~~~$ \{ $\eta'$.leftResolvedLiteral \} \;
        }
      }
       $\mathcal{S}(\eta)$ $\la$ $\bigcap_{\eta' \in \eta\textrm{.children}}$ safeLiteralsFrom($\eta'$)
    }
\caption{\label{algo:SetSafeLiterals} \texttt{setSafeLiterals}}
\end{footnotesize}
\end{algorithm}
\DecMargin{0.5em}

The set of safe literals of a node $\eta$ can be computed from the set of safe
literals of its children (cf.\ Algorithm~\ref{algo:SetSafeLiterals}). In the case when $\eta$ has a single child $\varsigma$, the safe literals of $\eta$ are simply the safe literals of $\varsigma$ together with the resolved literal $p$ of $\varsigma$ belonging to $\eta$ ($p$ is safe for $\eta$, because whenever $p$ is propagated down the proof through $\eta$, $p$ gets resolved in $\varsigma$). It is important to note, however, that if $\varsigma$ has been marked as regularized, it will eventually be replaced by $\eta$, and hence $p$ should not be added to the safe literals of $\eta$. In this case, the safe literals of $\eta$ should be exactly the same as the safe literals of $\varsigma$. When $\eta$ has several children, the safe literals of $\eta$ w.r.t. a child $\varsigma_i$ contain literals that are safe on all paths that go from $\eta$ through $\varsigma_i$ to the root. For a literal to be safe for all paths from $\eta$ to the root, it should therefore be in the intersection of the sets of safe literals w.r.t. each child.

The {\RP} and the {\RPI} algorithms differ from each other mainly in the
computation of the safe literals of a node that has many children. While {\RPI}
returns the intersection as shown in Algorithm~\ref{algo:SetSafeLiterals}, {\RP}
returns the empty set (cf.\ Algorithm~\ref{algo:SetSafeLiteralsRP}). Additionally, while in {\RPI} the safe literals of the root node contain all the literals of the root clause, in {\RP} the root node is always assigned an empty set of literals. 
(Of course, this makes a difference only when the proof is not a refutation.)
Note that during a traversal of the proof, 
the lines from 5 to 10 in Algorithm~\ref{algo:SetSafeLiterals} are executed as many times as the number of edges in the proof. 
Since every node has at most two parents, the number of edges is at most twice the number of nodes. 
Therefore, during a traversal of a proof with $n$ nodes, lines from 5 to 10 are
executed at most $2n$ times, and the algorithm remains linear.
In our prototype implementation, the sets of safe literals are instances of Scala's 
\texttt{mutable.HashSet} class. Being mutable, new elements can be added efficiently.
And being HashSets, membership checking is done in constant time in the average case, 
and set intersection (line 12) can be done in $O(k.s)$, where $k$ is the number of sets and $s$ is the size of the smallest set.\\

\IncMargin{0.5em}
\begin{algorithm}[p]
\begin{footnotesize}
\SetKwInOut{Input}{input}\SetKwInOut{Output}{output}
\SetKwData{units}{unitsQueue}
\SetKwData{fixedUnits}{fixedUnitsQueue}

\Input{A node $\eta$}
\Output{nothing (but the node $\eta$ gets a set of safe literals)}

\BlankLine

    \uIf{$\eta$ is a root node with no children}{
      $\mathcal{S}(\eta)$ $\la$ $\emptyset$ 
    }
    \Else{
      \uIf{$\eta$ has only one child $\eta'$}{
        \uIf{$\eta'$ is marked as regularized}{ 
          $\mathcal{S}(\eta)$ $\la$  $\mathcal{S}(\eta')$ \;}
        \uElseIf{$\eta$ is left parent of $\eta'$}{ 
        	$\mathcal{S}(\eta)$ $\la$  $\mathcal{S}(\eta')$ $\cup$ 
        	 \{ $\eta'$.rightResolvedLiteral \} \;
        }
        \ElseIf{$\eta$ is right parent of $\eta'$}{ 
			$\mathcal{S}(\eta)$ $\la$  $\mathcal{S}(\eta')$ $\cup$
			 \{ $\eta'$.leftResolvedLiteral \} \;
        }
      }
      \Else{
      	 $\mathcal{S}(\eta)$ $\la$ $\emptyset$
      }
    }
\caption{\label{algo:SetSafeLiteralsRP} \texttt{setSafeLiterals} for \RP}
\end{footnotesize}
\end{algorithm}
\DecMargin{0.5em}
%


\begin{example} 
\label{Example:Proof}
When applied to the proof $\psi$ shown in Figure \ref{ex:rpi-example-a}, the algorithm {\RPI} assigns $\{a,c\}$ and $\{a, \neg c\}$ as the safe literals of, respectively, $\eta_5$ and $\eta_8$. The safe literals of $\eta_4$ w.r.t. its children $\eta_5$ and $\eta_8$ are respectively $\{a,c,b\}$ and $\{a, \neg c, b\}$, and hence the safe literals of $\eta_4$ are $\{a,b\}$ (the intersection of $\{a,c,b\}$ and $\{a, \neg c, b\}$). Since the right resolved literal of $\eta_4$ ($a$) belongs to $\eta_4$'s safe literals, $\eta_4$ is correctly detected as a redundant node and hence regularized: $\eta_4$ is replaced by its right parent $\eta_3$. The resulting proof is shown in Figure \ref{ex:rpi-example-b}.



\end{example}

\section{Lifting to First-Order}\label{sec:Challenges}

In this section, we describe challenges that have to be overcome in order to successfully adapt {\RPI} to the first-order case. The first example illustrates the need to take unification into account. The other two examples discuss complex issues that can arise when unification is taken into account in a naive way.

\begin{example}\label{ex:unif} 
Consider the following proof $\psi$. When computed as in the propositional case, the safe literals for $\eta_3$ are $\{ Q(c), ~ P(a,x)\}$.

\begin{scriptsize}
\begin{prooftree}
\def\e{\mbox{\ $\vdash$\ }}
\AxiomC{$\eta_6$: $P(y,b)$ \e \hspace{-2cm}}
\AxiomC{$\eta_1$: \e $P(w,x)$}
\AxiomC{$\eta_2$: $P(w,x)$ \e $Q(c)$}
\BinaryInfC{$\eta_3$: \e $Q(c)$  \hspace{-1.5cm}}
\AxiomC{$\eta_4$: $Q(c)$ \e $P(a,x)$}
\BinaryInfC{$\eta_5$: \e $P(a,x)$}
\BinaryInfC{$\psi$: $\bot$}
\end{prooftree}
\end{scriptsize}

\noindent
As neither of $\eta_3$'s resolved literals is syntactically equal to a safe literal, the propositional {\RPI} algorithm would not change $\psi$. However, $\eta_3$'s left resolved literal $P(w,x)\in \eta_1$ is unifiable with the safe literal $P(a,x)$. Regularizing $\eta_3$, by deleting the edge between $\eta_2$ and $\eta_3$ and replacing $\eta_3$ by $\eta_1$, leads to further deletion of $\eta_4$ (because it is not resolvable with $\eta_1$) and finally to the much shorter proof below.

\begin{footnotesize}
\begin{prooftree}
\def\e{\mbox{\ $\vdash$\ }}
\AxiomC{$\eta_1$: \e$P(w,x)$}
\AxiomC{$\eta_6$: $P(y,b)$\e}
\BinaryInfC{$\psi'$: $\bot$}
\end{prooftree}
\end{footnotesize}

\end{example}

\noindent
Unlike in the propositional case, where a resolved literal must be syntactically equal to a safe literal for regularization to be possible, the example above suggests that, in the first-order case, it might suffice that the resolved literal be unifiable with a safe literal. However, there are cases, as shown in the example below, where mere unifiability is not enough and greater care is needed.

\begin{example}\label{ex:pairwise}

The node $\eta_3$ appears to be a candidate for regularization when the safe literals are computed as in the propositional case and unification is considered na\"{i}vely. Note that $\mathcal{S}(\eta_3)=\{Q(c), ~ P(a,x)\}$, and the resolved literal $P(a,c)$ is unifiable with the safe literal $P(a,x)$,

\begin{scriptsize}
\begin{prooftree}
\def\e{\mbox{\ $\vdash$\ }}
\AxiomC{$\eta_6$: $P(y,b)$ \e \hspace{-2cm}}
\AxiomC{$\eta_1$: \e $P(a,c)$}
\AxiomC{$\eta_2$: $P(a,c)$ \e $Q(c)$}
\BinaryInfC{$\eta_3$: \e $Q(c)$}
\AxiomC{$\eta_4$: $Q(c)$ \e $P(a,x)$}
\BinaryInfC{$\eta_5$: \e $P(a,x)$}
\BinaryInfC{$\psi$: $\bot$}
\end{prooftree}
\end{scriptsize}

\begin{figure*}
\begin{small}
\begin{prooftree}
\def\e{\mbox{\ $\vdash$\ }}
\AxiomC{$\eta_8$: $Q(f(a,e),c)\e$}
\AxiomC{$\eta_6$: $\e P(c,d)$ \hspace{-2cm}}
\AxiomC{$\eta_1$: $P(u,v)\e Q(f(a,v),u)$}
\AxiomC{$\eta_2$: $Q(f(a,x),y),Q(t,x)\e Q(f(a,z),y)$}
\BinaryInfC{$\eta_3$: $P(u,v),Q(t,v)\e Q(f(a,z),u)$}
\AxiomC{\hspace{-1cm} $\eta_4$: $\e Q(r,s)$}
\BinaryInfC{$\eta_5$: $P(u,v)\e Q(f(a,z),u)$}
\BinaryInfC{$\eta_7$: $\e Q(f(a,z),c)$}
\BinaryInfC{$\psi$: $\bot$}
\end{prooftree}
\end{small}
\caption{An example where pre-regularizability is not sufficient.}
\label{fig:ex-unifcheck}
\end{figure*}

\noindent
However, if we attempt to regularize the proof, the same series of actions as in Example \ref{ex:unif} would 
require resolution between $\eta_1$ and $\eta_6$, which is not possible.

\end{example}
One way to prevent the problem depicted above would be to require the resolved literal to be not only unifiable but subsume a safe literal. A weaker (and better) requirement is possible, and requires a slight modification of the concept of safe literals, taking into account the unifications that occur on the paths from a node to the root. 


\begin{definition}
The set of \emph{safe literals} for a node $\eta$ in a proof $\psi$ with root clause $\Gamma$, denoted $\mathcal{S}(\eta)$, is such that $\ell \in \mathcal{S}(\eta)$ if and only if $\ell \in \Gamma$ or for all paths from $\eta$ to the root of $\psi$ there is an edge $\n_1
\xrightarrow[\sigma]{\ell'} \n_2$ with $\ell' \sigma = \ell$.
\end{definition}

As in the propositional case, safe literals can be computed in a bottom-up traversal of the proof. Initially, at the root, the safe literals are exactly the literals that occur in the root clause. As we go up, the safe literals $\mathcal{S}(\eta')$ of a parent node $\eta'$ of $\eta$ where $\eta'
\xrightarrow[\sigma]{\ell} \eta$ is set to $\mathcal{S}(\eta) \cup \{ \ell \sigma \}$. Note that we apply the substitution to the resolved literal before adding it to the set of safe literals (cf. algorithm 3, lines 8 and 10). In other words, in the first-order case, the set of safe literals has to be a set of \emph{instantiated} resolved literals.

In the case of Example \ref{ex:pairwise}, computing safe literals as defined above would result in $\mathcal{S}(\eta_3)=\{Q(c),~P(a,b)\}$, where clearly the pivot $P(a,c)$ in $\eta_1$ is not safe. A generalization of this requirement is formalized below.

\begin{definition}
\label{prop:pair}
Let $\eta$ be a node with safe literals $\mathcal{S}(\eta)$ and parents $\eta_1$ and $\eta_2$, assuming without loss of generality, $\eta_1 \xrightarrow[\sigma_1]{\{\ell_1\} } \eta$.
The node $\eta$ is said to be \emph{pre-regularizable} in the proof $\psi$ if $\ell_1\sigma_1$ matches a safe literal $\ell^* \in \mathcal{S}(\eta)$.
\end{definition}

\noindent
This property states that a node is pre-regularizable if an instantiated resolved literal $\ell'$ matches a safe literal. The notion of \emph{pre-regulariziability} can be thought of as a \emph{necessary} condition for recycling the node $\eta$.

\begin{example}\label{ex:unifcheck}

Satisfying the pre-regularizability is not sufficient. Consider the proof $\psi$ in Figure \ref{fig:ex-unifcheck}. After collecting the safe literals, $\mathcal{S}(\eta_3) = \{\lnot Q(r,v),\lnot P(c,d), Q(f(a,e),c)\}$.
$\eta_3$'s pivot $Q(f(a,v),u)$ matches the safe literal $Q(f(a,e),c)$. Attempting to regularize $\eta_3$ would lead to the removal of $\eta_2$, the replacement of $\eta_3$ by $\eta_1$ and the removal of $\eta_4$ (because $\eta_1$ does not contain the pivot required by $\eta_5$), with $\eta_5$ also being replaced by $\eta_1$. Then resolution between $\eta_1$ and $\eta_6$ results in $\eta_7'$, which cannot be resolved with $\eta_8$, as shown below.

\begin{scriptsize}
\begin{prooftree}
\def\e{\mbox{\ $\vdash$\ }}
\AxiomC{$\eta_8$: $Q(f(a,e),c)\e$ \hspace{-0.5cm}}
\AxiomC{$\eta_6$: $\e P(c,d)$}
\AxiomC{$\eta_1$: $P(u,v)\e Q(f(a,v),u)$}
\BinaryInfC{$\eta_7'$: $\e Q(f(a,d),c)$}
\BinaryInfC{$\psi'$: ??}
\end{prooftree}
\end{scriptsize}

\noindent
$\eta_1$'s literal $Q(f(a, v), u)$, which would be resolved with $\eta_8$'s literal, was changed to $Q(f(a,d),c)$ due to the resolution between $\eta_1$ and $\eta_6$.
\end{example}


\noindent
Thus we additionally require that the following condition be satisfied.

\begin{definition} 
\label{prop:extracheck}
Let $\eta$ be pre-regularizable, with safe literals $\mathcal{S}(\eta)$ and parents $\eta_1$ and $\eta_2$, with clauses $\Gamma_1$ and $\Gamma_2$ respectively, assuming without loss of generality that $\eta_1 \xrightarrow[\sigma_1]{\{\ell_1\} } \eta$
such that $\ell_1\sigma_1$ matches a safe literal $\ell^*\in \mathcal{S}(\eta)$. 
The node $\eta$ is said to be \emph{strongly regularizable} in $\psi$ if $\Gamma_1 \sigma_{1} \sqsubseteq \mathcal{S}(\eta)$.
\end{definition}

This condition ensures that the remainder of the proof does not expect a variable in $\eta_1$ to be unified to different values simultaneously. This property is not necessary in the propositional case, as the literals of the replacement node would not change lower in the proof.


The notion of \emph{strongly regularizable} can be thought of as a \emph{sufficient} condition. 


\begin{thm}\label{thm:correct}
Let $\psi$ be a proof with root clause $\Gamma$ and $\eta$ be a node in $\psi$. Let $\psi^{\dagger} = \psi\setminus \{\eta\}$ and $\Gamma^{\dagger}$ be the root of $\psi^{\dagger}$. If $\eta$ is strongly regularizable, then $\Gamma^{\dagger} \sqsubseteq \Gamma$.
\end{thm}

\begin{proof} 
By definition of strong regularizability, $\eta$ is such
that there is a node $\eta'$ with clause $\Gamma'$ and such that
$\eta' \xrightarrow[\sigma']{\{\ell'\} } \eta$ and $\ell'\sigma'$
matches a safe literal $\ell^*\in \mathcal{S}(\eta)$ and
$\Gamma' \sigma' \sqsubseteq \mathcal{S}(\eta)$.

Firstly, in $\psi^{\dagger}$, $\eta$ has been replaced by $\eta'$. Since
$\Gamma' \sigma' \sqsubseteq \mathcal{S}(\eta)$, by definition of
$\mathcal{S}(\eta)$, every literal $\ell$ in $\Gamma'$ either subsumes 
a single literal that occurs as a pivot on every path
from $\eta$ to the root in $\psi$ (and hence on every new path from
$\eta'$ to the root in $\psi^{\dagger}$) or subsumes literals 
$\ell \sigma_1$,\ldots,$\ell\sigma_n$ in $\Gamma$. In the former case,
$\ell$ is resolved away in the construction of $\psi^{\dagger}$ (by
contracting the descendants of $\ell$ with the pivots in each path).
In the latter case, the literal $\ell \sigma_k$ ($1 \leq k \leq n$) in
$\Gamma$ is a descendant of $\ell$ through a path $k$ and the
substitution $\sigma_k$ is the composition of all substitutions on
this path. When $\eta$ is replaced by $\eta'$, two things may happen
to $\ell \sigma_k$. If the path $k$ does not go through $\eta$, 
$\ell \sigma_k$ remains unchanged (i.e. $\ell \sigma_k \in \Gamma^{\dagger}$
unless the path $k$ ceases to exist in $\psi^{\dagger}$). If the path
$k$ goes through $\eta$, the literal is changed to 
$\ell\sigma^{\dagger}_k$, where $\sigma^{\dagger}_k$ is such that 
$\sigma_k = \sigma' \sigma^{\dagger}_k$.

Secondly, when $\eta$ is replaced by $\eta'$, the edge from
$\eta$'s other parent $\eta''$ to $\eta$ ceases to exist in
$\psi^{\dagger}$. Consequently, any literal $\ell$ in $\Gamma$ that is a
descendant of a literal $\ell''$ in the clause of $\eta''$ through a
path via $\eta$ will not belong to $\Gamma^{\dagger}$.


Thirdly, a literal from $\Gamma$ that descends neither from $\eta'$ nor from $\eta''$ either remains unchanged in $\Gamma^{\dagger}$ or, if the path to the node from which it descends ceases to exist in the construction of $\psi^{\dagger}$, does not belong to $\Gamma^{\dagger}$ at all.

Therefore, by the three facts above, $\Gamma^{\dagger} \sigma' \sqsubseteq \Gamma$, and hence $\Gamma^{\dagger} \sqsubseteq \Gamma$.
\end{proof}

As the name suggests, strong regularizability is stronger than necessary. In some cases, nodes may be regularizable even if they are not strongly regularizable. A weaker condition (conjectured to be sufficient) is presented below. This alternative relies on knowledge of how literals are changed after the deletion of a node in a proof (and it is inspired by the \emph{post-deletion unifiability condition} described for {\FOLowerUnits} in \cite{GFOLU}). However, since weak regularizability is more complicated to check, it is not as suitable for implementation as strong regularizability. 
\begin{definition}\label{def:postdelprop}
Let $\eta$ be a pre-regularizable node with parents $\eta_1$ and $\eta_2$, assuming without loss of generality that $\eta_1 \xrightarrow[\sigma_1]{\{\ell_1\} } \eta$ 
such that $\ell_1$ is unifiable with some $\ell^* \in \mathcal{S}(\eta)$.
For each safe literal $\ell = \ell_s\sigma_s \in \mathcal{S}(\eta_1)$, let $\eta_\ell$ be a node on the path from $\eta$ to the root of the proof such that $\abs{\ell}$ is the pivot of $\eta_\ell$.
Let $\mathcal{R}(\eta_\ell)$ be the set of all resolved literals $\ell_s'$ such that $\eta_2' \xrightarrow[\sigma_s]{\{\ell_s\} } \eta_\ell$, $\eta_1' \xrightarrow[\sigma_s']{\{\ell_s'\} } \eta_\ell$, and $\ell_s\sigma_s=\dual{\ell_s'}\sigma_s'$, for some nodes $\eta_2'$ and $\eta_1'$ and unifier $\sigma_s'$; if no such node $\eta_\ell$ exists, define $\mathcal{R}(\eta_\ell)=\emptyset$.
The node $\eta$ is said to be \emph{weakly regularizable} in $\psi$ if, for all $\ell \in \mathcal{S}(\eta_1)$, all elements in $\mathcal{R}^{\dagger}(\eta_\ell) \cup \{ \dual{\ell}^\dagger \}$ are unifiable, where $\dual{\ell}^{\dagger}$ is the literal in $\dn{\psi}{\eta_2}$ that used to be\footnote{Because of the removal of $\eta_2$, $\dual{\ell}^{\dagger}$ may differ from $\dual{\ell}$.} $\dual{\ell}$ in $\psi$ and $\mathcal{R}^{\dagger}(\eta_\ell)$ is the set of literals in $\dn{\psi}{\eta_2}$ that used to be the literals of $\mathcal{R}(\eta_\ell)$ in $\psi$.
\end{definition}


This condition requires the ability to determine the underlying (uninstantiated) literal for each safe literal of a weakly regularizable node $\eta$. To achieve this, one could store safe literals as a pair $(\ell_s,\sigma_s)$, rather than as an instantiated literal $\ell_s\sigma_s$, although this is not necessary for the previous conditions.

Note further that there is always at least one node $\eta_\ell$ as assumed in the definition for any safe literal which was not contained in the root clause of the proof: the node which resulted in $\ell = \ell_s\sigma_s \in \mathcal{S}(\eta)$ being a safe literal for the path from $\eta$ to the root of the proof. Furthermore, it does not matter which node $\eta_\ell$ is used. To see this, consider some node $\eta_\ell' \neq \eta_\ell$ with the same pivot $\abs{\ell}=\abs{\ell_s\sigma_s}$. Consider arbitrary nodes $\eta_1$ and $\eta_2$ such that  $\eta_2 \xrightarrow[\sigma_s]{\{\ell_s\} } \eta_\ell$ and $\eta_1 \xrightarrow[\sigma_1]{\{\ell_1\} } \eta_\ell$ where $\ell_s\sigma_s=\dual{\ell_1}\sigma_1$. Now consider arbitrary nodes $\eta_1'$ and $\eta_2'$ such that  $\eta_2' \xrightarrow[\sigma_s]{\{\ell_s\} } \eta_\ell'$ and $\eta_1' \xrightarrow[\sigma_1']{\{\ell_1'\} } \eta_\ell'$ where $\ell_s\sigma_s=\dual{\ell_1'}\sigma_1'$. Since the pivots for $\eta_\ell$ and $\eta_\ell'$ are equal, we must have that 
$\abs{\ell_s\sigma_s}=\abs{\ell_1\sigma_1}$ and $\abs{\ell_s\sigma_s}=\abs{\ell_1'\sigma_1'}$, and thus $\abs{\ell_1\sigma_1}=\abs{\ell_1'\sigma_1'}$. This shows that it does not matter which $\eta_\ell$ we use; the instantiated resolved literals will always be equal implying that both of the resolved literals $\ell_1$ and $\ell_1'$ will be contained in both $\mathcal{R}(\eta_\ell)$ and $\mathcal{R}(\eta_\ell')$.

Informally, a node $\eta$ is weakly regularizable in a proof if it can be replaced by one of its parents $\eta_1$, such that for each $\ell \in \mathcal{S}(\eta_1)$, $\abs{\ell}$ can still be used as a pivot in order to complete the proof. Weakly regularizable nodes differ from strongly regularizable nodes by not requiring the entire parent $\eta_1$ replacing the resolution $\eta$ to be simultaneously matched to a subset of $\mathcal{S}(\eta)$, and requires knowledge of how literals will be instantiated after the removal of $\eta_2$ and $\eta$ from the proof.


\begin{table}
\centering
\begin{tabular}{| c | c | c | c | }
\hline
$\eta$ & $\mathcal{S}(\eta)$ & $\mathcal{R}(\eta)$ & $\mathcal{R}^\dagger(\eta)$ \\ \hline \hline
$\eta_1$ &  $\{P(w)\}$ & $\emptyset$  & $\emptyset$\\ \hline 
$\eta_2$ &  $\{\lnot P(w)\}$ & $\emptyset$  & $\emptyset$\\ \hline 
$\eta_3$ &  $\{R(a),\lnot P(w)\}$ & $\emptyset$  & $\emptyset$\\ \hline 
$\eta_4$ &  $\{\lnot R(a),\lnot P(w)\}$& $\emptyset$& $\emptyset$ \\ \hline 
$\eta_5$ &  $\{Q(z),\lnot R(a), \lnot P(w)\}$ & $\emptyset$ & $\emptyset$\\ \hline 
$\eta_6$ &  $\{\lnot P(w), \lnot Q(z), \lnot R(a) \}$ & $\{P(u),P(y)\}$& $\{P(u)\}$\\ \hline 
$\eta_7$ &  $\{P(y), \lnot P(w), \lnot Q(z), \lnot R(a) \}$ & $\emptyset$ & $\emptyset$ \\ \hline 
$\eta_8$ &   $\{\lnot P(y), \lnot P(w), \lnot Q(z), \lnot R(a) \}$ & $\emptyset$ & $\emptyset$\\ \hline 
\end{tabular}
\hfill
\caption{The sets $\mathcal{S}(\eta)$ and $\mathcal{R}(\eta)$ for each node $\eta$ in the first proof of Example \ref{ex:weak}.}
\label{tab:exweakreg}
\end{table}

\begin{example}\label{ex:weak}
This example illustrates a case where a node is weakly regularizable but not strongly regularizable. Table \ref{tab:exweakreg} shows the sets $\mathcal{S}(\eta)$, $\mathcal{R}(\eta)$ and $\mathcal{R}^\dagger(\eta)$ for the nodes $\eta$ in the proof below. Observe that $\eta_6$ is pre-regularizable, since $\lnot P(x)$ is unifiable with $\lnot P(w)\in \mathcal{S}(\eta_6)$. In fact, $\eta_6$ is the only pre-regularizable node in the proof, and thus the sets $\mathcal{R}(\eta) = \emptyset$ for all $\eta \neq \eta_6$.
In the proof below, note that $\eta_6$ is not strongly regularizable: there is no unifier $\sigma$ such that $\{\lnot P(x),\lnot Q(x),\lnot R(x)\} \sigma \subseteq \mathcal{S}(\eta_6)$.
\begin{scriptsize}
\begin{prooftree}
\def\e{\mbox{\ $\vdash$\ }}
\AxiomC{$\eta_1$: $\e P(u)$ \hspace{-2cm}}
\AxiomC{$\eta_5$: $P(z) \e Q(z)$ \hspace{-0.5cm}}
\AxiomC{$\eta_8$: $P(x),Q(x),R(a)\e$}
\AxiomC{$\eta_7$: $\e P(y)$  \hspace{-1cm}}
\BinaryInfC{$\eta_6$: $Q(y),R(a)\e$ }
\BinaryInfC{$\eta_4$: $P(z),R(a)\e$ \hspace{-2cm} }
\AxiomC{ \hspace{-1cm} $\eta_3$: $\e R(a)$}
\BinaryInfC{ $\eta_2$: $P(z)\e$}

\BinaryInfC{$\psi$: $\bot$}
\end{prooftree}

\end{scriptsize}
\noindent
We show that $\eta_6$ is weakly regularizable, and that $\eta_7$ can be removed. Recalling that $\eta_6$ is pre-regularizable, observe that $\mathcal{R}^\dagger(\eta_6) \cup \{\dual{\lnot P(w)}\}$ is unifiable.
Consider the following proof of $\psi \setminus \{\eta_7\}$:
\begin{scriptsize}
\begin{prooftree}
\def\e{\mbox{\ $\vdash$\ }}
\AxiomC{$\eta_1$: $\e P(u)$ \hspace{-1.75cm}}
\AxiomC{$\eta_8$: $P(x),Q(x),R(a)\e$}
\AxiomC{$\eta_5$: $P(z) \e Q(z)$}
\BinaryInfC{$\eta_4'$: $P(z), P(z),R(a)\e$}
\UnaryInfC{$\eta_4$: $P(z),R(a)\e$}
\AxiomC{$\eta_3$: $\e R(a)$}
\BinaryInfC{$\eta_2$: $P(z)\e$}
\BinaryInfC{$\psi$: $\bot$}
\end{prooftree}
\end{scriptsize}
Now observe that for each $\ell \in \mathcal{S}(\eta_8)$ we have the following, showing that $\eta_6$ is weakly regularizable:
\begin{itemize}
\item $\ell=\lnot  Q(y)$: $\ell^\dagger = \lnot Q(x)$ which is unifiable with $\dual{\ell}^\dagger=Q(z)$
\item $\ell=\lnot R(a)$: $\ell^\dagger = \lnot R(a)$ which is (trivially) unifiable with $\dual{\ell}^\dagger=R(a)$
\item $\ell=\lnot P(w)$: $\ell^\dagger = \lnot P(z)$ which is unifiable with $\dual{\ell}^\dagger=P(u)$
\item $\ell=\lnot P(y)$: $\ell^\dagger = \lnot P(z)$ which is unifiable with $\dual{\ell}^\dagger=P(u)$
\end{itemize}
\end{example}

If a node $\eta$ with parents $\eta_1$ and $\eta_2$ is pre-regularizable and strongly regularizable in $\psi$, then $\eta$ is also weakly regularizable in $\psi$.

\section{Implementation}
\label{sec:FORPI}
{\FirstOrderRPI} ({\FORPI}) (cf. Algorithm \ref{algo:FORPI}) is a first-order generalization of the propositional {\RPI}.
{\FORPI} traverses the proof in a bottom-up manner, storing for every node a set of safe literals. The set of safe literals for a node $\psi$ is computed from the set of safe literals of its children (cf.\ Algorithm~\ref{algo:foSetSafeLiterals}), similarly to the propositional case, but additionally applying unifiers to the resolved literals (cf. Example \ref{ex:pairwise}).
If one of the node's resolved literals matches a literal in the set of safe literals, then it may be possible to regularize the node by replacing it by one of its parents.

\IncMargin{0.5em}
\begin{algorithm}[bt]
\begin{footnotesize}
\SetKwInOut{Input}{input}\SetKwInOut{Output}{output}
\SetKwData{units}{unitsQueue}
\SetKwData{fixedUnits}{fixedUnitsQueue}

\Input{A first-order proof $\psi$}
\Output{A possibly less-irregular first-order proof $\psi'$}

\BlankLine

$\psi'$ $\la$ $\psi$\;
traverse $\psi'$ bottom-up and \ForEach{node $\eta$ in $\psi'$}{
   \If{$\eta$ is a resolvent node}{
     setSafeLiterals($\eta$) \;
     regularizeIfPossible($\eta$)
   }
  }
$\psi'$ $\la$ fix($\psi'$) \;
\Return {$\psi'$}\;
\caption{\label{algo:FORPI} \texttt{\FORPI}}
\end{footnotesize}
\end{algorithm}
\DecMargin{0.5em}    


In the first-order case, we additionally check for strong regularizability (cf. lines 2 and 6 of Algorithm~\ref{algo:foRegularize}).
Similarly to {\RPI}, instead of replacing the irregular node by one of its parents immediately, 
its other parent is marked as a \texttt{deletedNode}, as shown in Algorithm~\ref{algo:foRegularize}.
As in the propositional case, fixing of the proof is postponed to another (single) traversal, as regularization proceeds top-down and only nodes below a regularized node may require fixing.
During fixing, the irregular node is actually replaced by the parent that is not marked as \texttt{deletedNode}. During proof fixing, factoring inferences can be applied, in order to compress the proof further.

Note that, in order to reduce notation clutter in the pseudocodes, we slightly abuse notation and do not explicitly distinguish proofs, their root nodes and the clauses stored in their root nodes. It is clear from the context whether $\psi$ refers to a proof, to its root node or to its root clause.

\IncMargin{0.5em}
\begin{algorithm}[bt]
\begin{footnotesize}

\SetKwInOut{Input}{input}\SetKwInOut{Output}{output}
\SetKwData{units}{unitsQueue}
\SetKwData{fixedUnits}{fixedUnitsQueue}

\Input{A node $\psi=\psi_L  \res{\ell_L}{\sigma_L}{\ell_R}{\sigma_R} \psi_R$}
\Output{nothing (but the proof containing $\psi$ may be changed)}

\BlankLine
    \uIf{$\exists \sigma$  and $\ell \in \mathcal{S}(\psi)$ such that $\ell = \ell_R\sigma_R\sigma$}{
     \uIf{$\psi_R\sigma_R\sigma \subseteq \mathcal{S}(\psi)$} {
      mark $\psi_L$ as \texttt{deletedNode} \;
      mark $\psi$ as regularized
}
    }
    \ElseIf{$\exists \sigma$  and $\ell \in \mathcal{S}(\psi)$ such that $\ell = \ell_L\sigma_L\sigma$ }{
     \uIf{$\psi_L\sigma_L\sigma \subseteq\mathcal{S}(\psi)$} {
      mark $\psi_R$ as \texttt{deletedNode} \;
      mark $\psi$ as regularized
}
    }
\caption{\label{algo:foRegularize} \texttt{regularizeIfPossible} for \texttt{FORPI}}
\end{footnotesize}
\end{algorithm}
\DecMargin{0.5em}    

\IncMargin{0.5em}
\begin{algorithm}[bt]
\begin{footnotesize}

\SetKwInOut{Input}{input}\SetKwInOut{Output}{output}
\SetKwData{units}{unitsQueue}
\SetKwData{fixedUnits}{fixedUnitsQueue}

\Input{A first-order resolution node $\psi$}
\Output{nothing (but the node $\psi$ gets a set of safe literals)}

\BlankLine

    \uIf{$\psi$ is a root node with no children}{
      $\mathcal{S}(\psi) \la$ $\psi$.clause  
    }
    \Else{
      \ForEach{$\psi'$ $\in$ $\psi${\upshape.children}}{
        \uIf{$\psi'$ is marked as regularized}{ 
          safeLiteralsFrom($\psi'$) $\la$ $\mathcal{S}(\psi')$ \;}
          \uElseIf{$\psi' = \psi  \res{\ell_L}{\sigma_L}{\ell_R}{\sigma_R} \psi_R$ for some $\psi_R$}{ 
        	safeLiteralsFrom($\psi'$) $\la$ $\mathcal{S}(\psi')~ \cup $ $\{\ell_R\sigma_R  \}$ 
        }
        \ElseIf{$\psi' = \psi_L  \res{\ell_L}{\sigma_L}{\ell_R}{\sigma_R} \psi$ for some $\psi_L$}{ 
	safeLiteralsFrom($\psi'$) $\la$ $\mathcal{S}(\psi') ~\cup $ $\{ \ell_L\sigma_L \}$
        }
      }
      $\mathcal{S}(\psi)$ $\la$ $\bigcap_{\psi' \in \psi\textrm{.children}}$ safeLiteralsFrom($\psi'$)
    }

\caption{\label{algo:foSetSafeLiterals} \texttt{setSafeLiterals} for \texttt{FORPI}}
\end{footnotesize}
\end{algorithm}
\DecMargin{0.5em}

\section{Experiments} \label{sec:exp}

A prototype version of {\FORPI} has been implemented in the functional programming language Scala as part of the \skeptik
library. This library includes an implementation of {\GFOLU} \cite{GFOLU}. In order to evaluate the algorithm's effectiveness, {\FORPI} was tested on two data sets: proofs generated by a real theorem prover and randomly-generated resolution proofs. The proofs are included in the source code repository, available at \url{https://github.com/jgorzny/Skeptik}. Note that by implementing the algorithms in this library, we have a relative guarantee that the compressed proofs are correct, as in \skeptik every inference rule (e.g. resolution, factoring) is implemented as a small class (each at most 178 lines of code that is assumed correct) with a constructor that checks whether the conditions for the application of the rule are met, thereby preventing the creation of objects representing incorrect proof nodes (i.e. unsound inferences). We only need to check that the root clause of the compressed proof is equal to or stronger than the root clause of the input proof and that the set of axioms used in the compressed proof is a (possibly non-proper) subset of the set of axioms used in the input proof.



First, {\FORPI} was evaluated on the same proofs used to evaluate {\GFOLU}. These proofs were generated by executing the {\SPASS} theorem prover ({\url{http://www.spass-prover.org/}) on 1032 real-world unsatisfiable first-order problems without equality from the TPTP Problem Library \cite{TPTP}. In order to generate pure resolution proofs, the advanced inference rules of {\SPASS} were disabled: the only enabled inference rules used were ``Standard Resolution'' and ``Condensation''. The proofs were originally generated on the Euler Cluster at the University of Victoria with a time limit of 300 seconds per problem. Under these conditions, {\SPASS} was able to generate 308 proofs. The proofs generated by {\SPASS} were small: proof lengths varied from 3 to 49, and the number of resolutions in a proof ranged from 1 to 32.


In order to test {\FORPI}'s effectiveness on larger proofs, a total of 2280 proofs were randomly generated and then used as a second benchmark set. The randomly generated proofs were much larger than those of the first data set: proof lengths varied from 95 to 700, while the number of resolutions in a proof ranged from 48 to 368.

\subsection{Proof Generation}
Additional proofs were generated by the following procedure: start with a root node whose conclusion is $\bot$, and make two premises $\eta_1$ and $\eta_2$ using a randomly generated literal such that the desired conclusion is the result of resolving $\eta_1$ and $\eta_2$. For each node $\eta_i$, determine the inference rule used to make its conclusion: with probability $p=0.9$, $\eta_i$ is the result of a resolution, otherwise it is the result of  factoring. 

Literals are generated by uniformly choosing a number from $\{1,\dots,k,k+1\}$ where $k$ is the number of predicates generated so far; if the chosen number $j$ is between $1$ and $k$, the $j$-th predicate is used; otherwise, if the chosen number is $k+1$, a new predicate with a new random arity (at most four) is generated and used. Each argument is a constant with probability $p=0.7$ and a complex term (i.e. a function applied to other terms) otherwise; functions are generated similarly to predicates. 

If a node $\eta$ should be the result of a resolution, then with probability $p=0.2$ we generate a left parent $\eta_\ell$ and a right parent $\eta_r$ for $\eta$ (i.e. $\eta = \eta_\ell \odot \eta_r$) having a common parent $\eta_c$ (i.e. $\eta_l = (\eta_\ell)_\ell \odot \eta_c$ and $\eta_r = \eta_c \odot (\eta_r)_r$, for some newly generated nodes $(\eta_\ell)_\ell$ and $(\eta_r)_r$ ). The common parent ensures that also non-tree-like DAG proofs are generated. 

This procedure is recursively applied to the generated parent nodes. 
Each parent of a resolution has each of its terms not contained in the pivot replaced by a fresh variable with probability $p=0.7$.
At each recursive call, the additional minimum height required for the remainder of the branch is decreased by one with probability $p=0.5$. Thus if each branch always decreases the additional required height, the proof has height equal to the initial minimum value. The process stops when every branch is required to add a subproof of height zero or after a timeout is reached. In any case, the topmost generated node for each branch is generated as an axiom node. 

The minimum height was set to 7 (which is the minimum number of nodes in an irregular proof plus one) and the timeout was set to 300 seconds (the same timeout allowed for {\SPASS}). The probability values used in the random generation were carefully chosen to produce random proofs similar in shape to the real proofs obtained by {\SPASS}. For instance, the probability of a new node being a resolution (respectively, factoring) is approximately the same as the frequency of resolutions (respectively, factorings) observed in the real proofs produced by {\SPASS}.

\subsection{Results}
For consistency, the same system and metrics were used. Proof compression and proof generation was performed on a laptop (2.8GHz Intel Core i7 processor with 4GB of RAM (1333MHz DDR3) available to the Java Virtual Machine). For each proof $\psi$, we measured the time needed to compress the proof ($t(\psi)$) and the compression ratio ($(|\psi|-|\alpha(\psi)|)/|\psi|$) where $|\psi|$ is the number of resolutions in the proof, and $\alpha(\psi)$ is the result of applying a compression algorithm or some composition of {\FORPI} and {\GFOLU}. Note that we consider only the number of resolutions in order to compare the results of these algorithms to their propositional variants (where factoring is implicit). Moreover, factoring could be made implicit within resolution inferences even in the first-order case and we use explicit factoring only for technical convenience.


\begin{table*}[bt]
\hspace*{-1.5cm}
\begin{tabular}{| l || r | r | r || r | r | r  | }
\hline
 Algorithm& \multicolumn{3}{c ||}{\# of Proofs Compressed} & \multicolumn{3}{c |}{\# of Removed Nodes}  \\
& \multicolumn{1}{c }{TPTP} & \multicolumn{1}{c}{Random}  & \multicolumn{1}{c ||}{Both}  & \multicolumn{1}{c }{TPTP} & \multicolumn{1}{c }{Random} & \multicolumn{1}{c |}{Both}  \\ \hline \hline
{\GFOLU}(p) & 55 (17.9\%) & 817 (35.9\%) & 872 (33.7\%)  & 107 (4.8\%) & 17,769 (4.5\%) & 17,876 (4.5\%)    \\ \hline
{\FORPI}(p)  & 23 (7.5\%) &  666 (29.2\%) & 689 (26.2\%)  &  36 (1.6\%) &  28,904 (7.3\%) &  28,940 (7.3\%)   \\ \hline
{\GFOLU}({\FORPI}(p))   & 55 (17.9\%) & 1303 (57.1\%) & 1358 (52.5\%) & 120 (5.4\%)  & 48,126 (12.2\%) & 48,246 (12.2\%) \\ \hline
{\FORPI}({\GFOLU}(p)) & 23 (7.5\%) & 1302  (57.1\%)&  1325 (51.2\%) & 120 (5.4\%) & 48,434 (12.3\%) & 48,554 (12.3\%)  \\ \hline
Best                            & 59 (19.2\%) & 1303 (57.1\%) & 1362 (52.5\%)   & 120 (5.4\%) & 55,530 (14.1\%) & 55,650 (14.0\%)     \\ \hline
\end{tabular} $~~~~~~~~~~~~~~~~$
\vspace{5pt}
\caption{Number of proofs compressed and number of overall nodes removed}
\label{tab:results}
\end{table*}

\begin{table*}[bt]
\centering
\begin{tabular}{| l | r | r || l | r |}
\hline
Algorithm &  \multicolumn{2}{c ||}{First-Order Compression}  &  Algorithm & Propositional Compression \cite{Boudou}  \\
& $~~~~$ All   & Compressed Only & & \\ \hline \hline
{\GFOLU}(p) &  4.5\%& 13.5\% &{\LU}(p) & 7.5\% \\ \hline
{\FORPI}(p) & 6.2\%&  23.2\%&{\RPI}(p) &  17.8\% \\ \hline
{\GFOLU}({\FORPI}(p)) &  10.6\%& 23.0\%& ({\LU}({\RPI}(p)) &  21.7\% \\ \hline
{\FORPI}({\GFOLU}(p)) &  11.1\%& 21.5\%& ({\RPI}({\LU}(p)) & 22.0\% \\ \hline
Best & 12.6\% & 24.4\%&  Best &  22.0\% \\ \hline
\end{tabular}
\vspace{5pt}
\caption{Mean compression results}
\label{tab:result-mean}
\end{table*}

Table \ref{tab:results} summarizes the results of {\FORPI} and its combinations with {\GFOLU}. The first set of columns describes the percentage of proofs that were compressed by each compression algorithm. The algorithm `Best' runs both combinations of {\GFOLU} and {\FORPI} and returns the shortest proof output by either of them. The total number of proofs is $308+2280=2588$ and the total number of resolution nodes is $2,249 + 393,883
= 396,132$. The percentages in the last three columns are computed by $(\Sigma_{\psi \in \Psi} |\psi|  - \Sigma_{\psi\in \Psi} |\alpha(\psi)|)/(\Sigma_{\psi \in \Psi} |\psi|)$ for each data set $\Psi$ (TPTP, Random, or Both). The use of {\FORPI} alongside {\GFOLU} allows at least an additional 17.5\% of proofs to be compressed. Furthermore, the use of both algorithms removes almost twice as many nodes than any single algorithm.

Table \ref{tab:result-mean} compares the results of {\FORPI} and its combinations with {\GFOLU} with their propositional variants as evaluated in  \cite{Boudou}. The first column describes the mean compression ratio for each algorithm including proofs that were not compressed by the algorithm, while the second column calculates the mean compression ratio considering only compressed proofs. It is unsurprising that the first column is lower than the propositional mean for each algorithm: there are stricter requirements to apply these algorithms to first-order proofs. In particular, additional properties must be satisfied before a unit can be lowered, or before a pivot can be recycled. On the other hand, when first-order proofs are compressed, the compression ratios are on par with or better than their propositional counterparts.

\begin{figure*}[p]
    
  \subfloat[Number of (non-)compressed proofs]{{\includegraphics[scale=0.5]{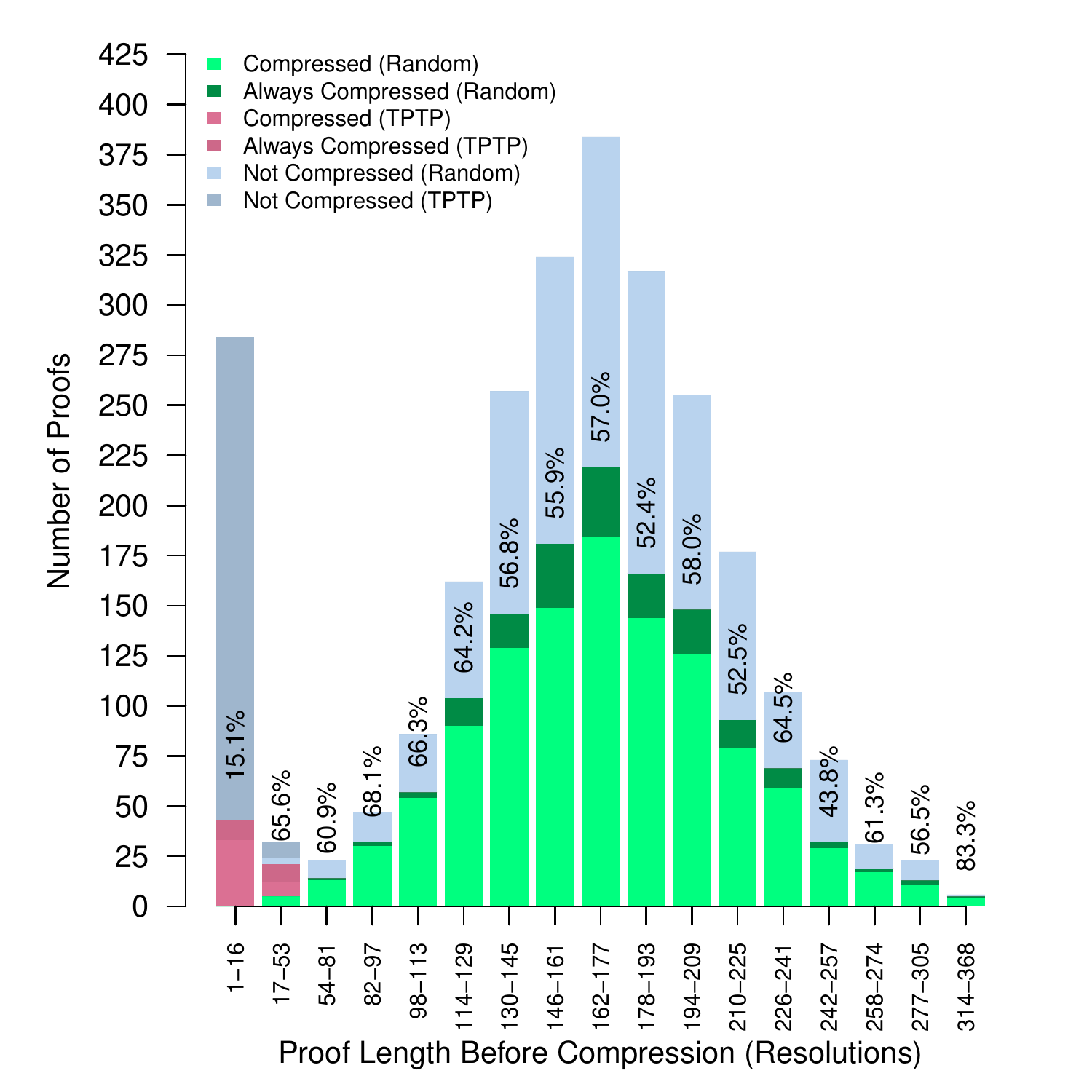} }}
   \subfloat[Compressed length against input length]{{\includegraphics[scale=0.5]{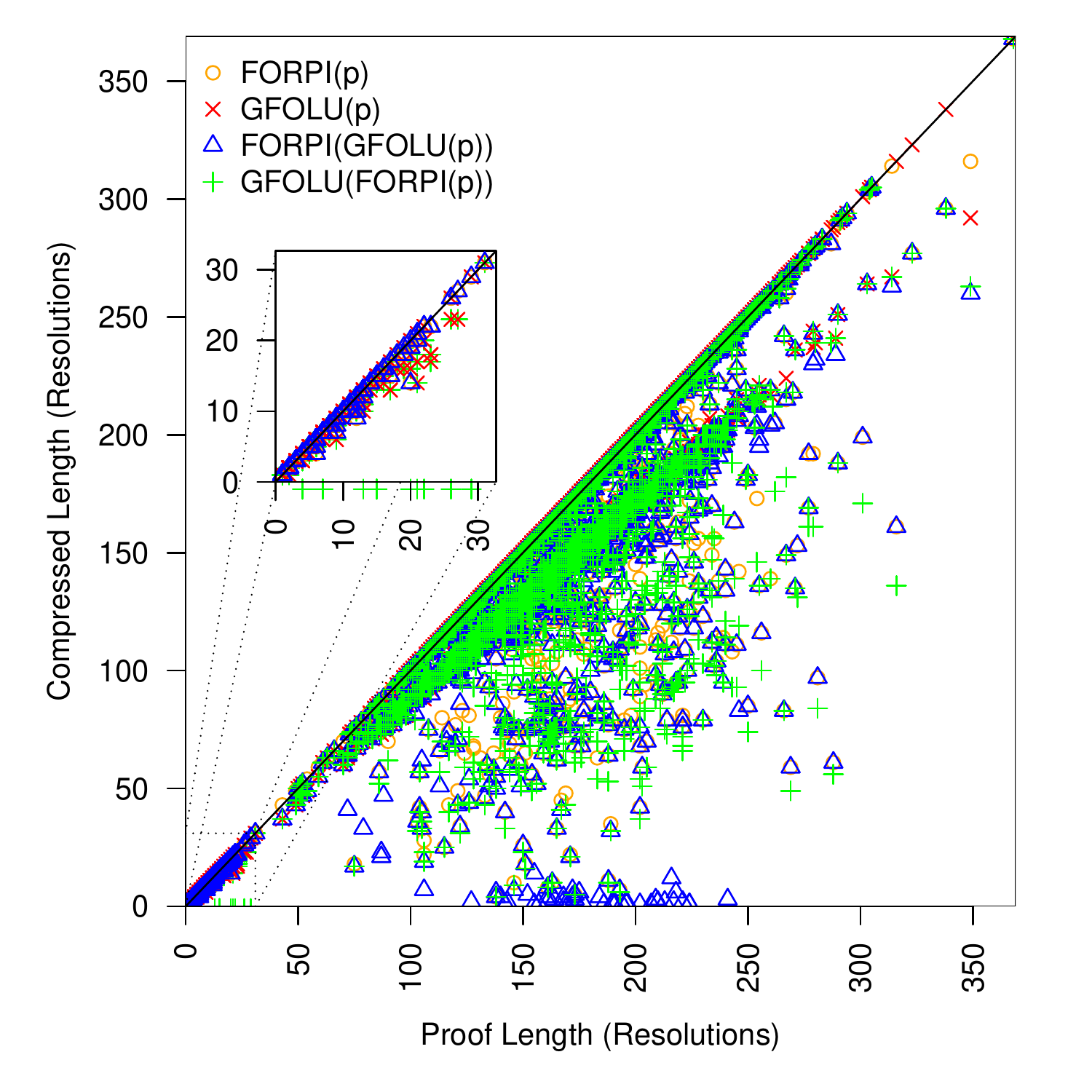} }}   
   
\subfloat[\FORPI(\GFOLU(p)) vs. \GFOLU(\FORPI(p))]{{\includegraphics[scale=0.5]{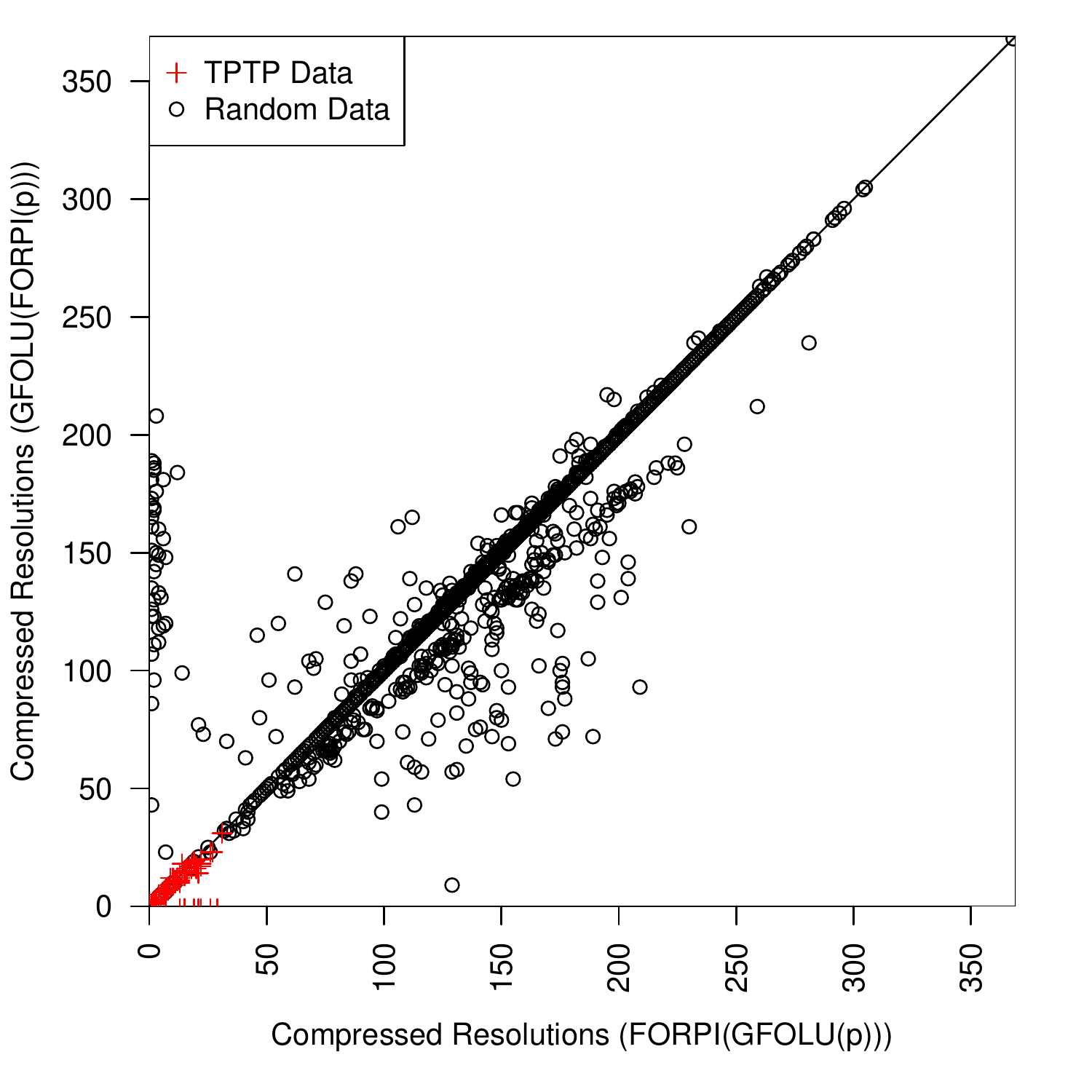} }}
  \subfloat[Cumulative proof compression]{{\includegraphics[scale=0.5]{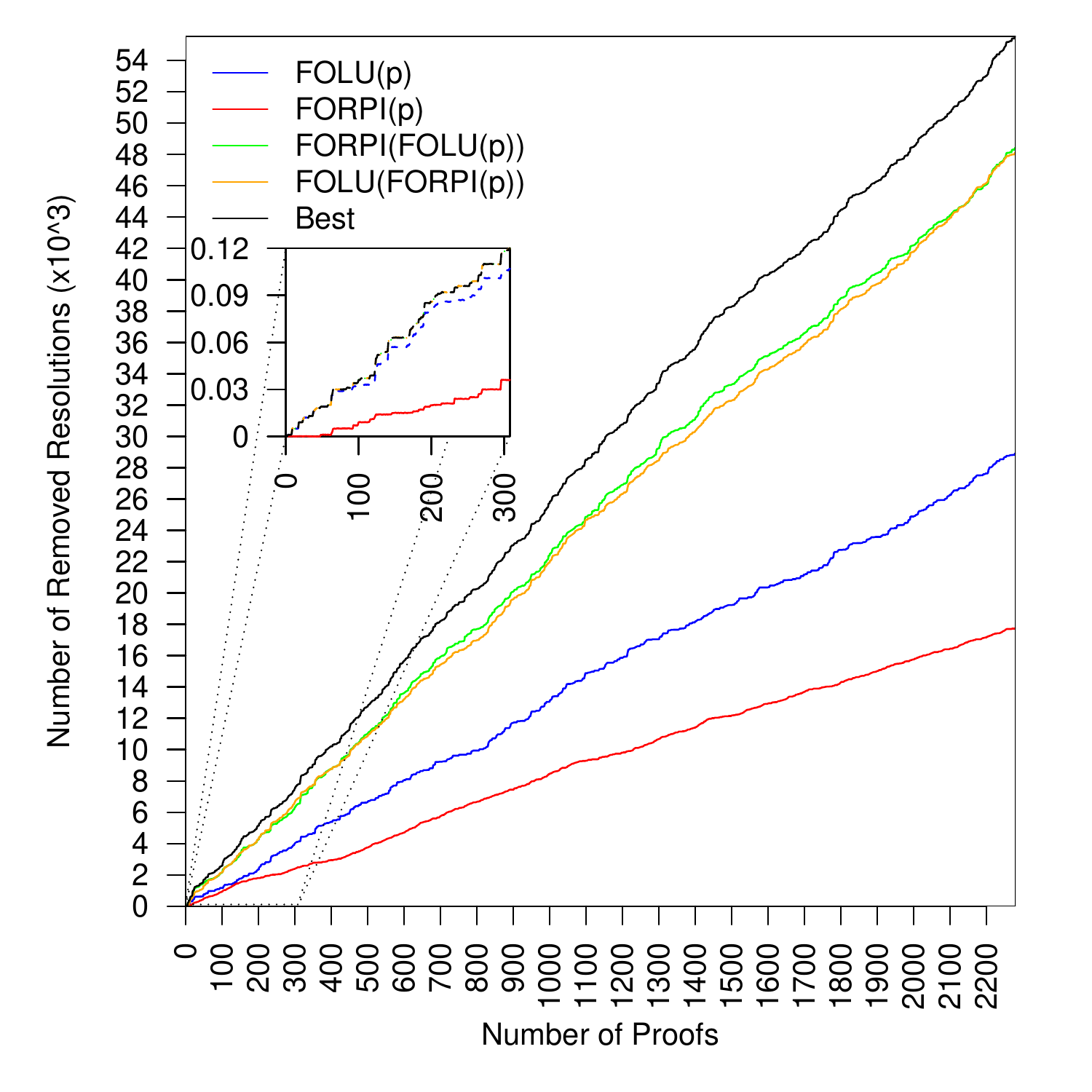} }}

 \caption{\GFOLU \& \FORPI Combination Results}
\label{fig:ex}

\end{figure*}

Figure \ref{fig:ex} (a) shows the number of proofs (compressed and uncompressed) per grouping based on number of resolutions in the proof. The red (resp. dark grey) data shows the number of compressed (resp. uncompressed) proofs for the TPTP data set, while the green (resp. light grey) data shows the number of compressed (resp. uncompressed) proofs for the random proofs. The number of proofs in each group is the sum of the heights of each coloured bar in that group. The overall percentage of proofs compressed in a group is indicated on each bar. Dark colors indicate the number of proofs compressed by {\FORPI}, {\GFOLU}, and both compositions of these algorithms; light colors indicate cases were {\FORPI} succeeded, but at least one of {\GFOLU} or a combination of these algorithms achieved zero compression. 
Given the size of the TPTP proofs, it is unsurprising that few are compressed: small proofs are a priori less likely to contain irregularities. On the other hand, 
at least 43\% of the randomly generated proofs in each size group could be compressed.


Figure \ref{fig:ex} (b) is a scatter plot comparing the number of resolutions of the input proof against the number of resolutions in the compressed proof for each algorithm. The results on the TPTP data are magnified in the sub-plot. For the randomly generated proofs (points outside of the sub-plot), it is often the case that the compressed proof is significantly shorter than the input proof. Interestingly, {\GFOLU} appears to reduce the number of resolutions by a linear factor in many cases. This is likely due to a linear growth in the number of non-interacting irregularities (i.e. irregularities for which the lowered units share no common literals with any other sub-proofs), which leads to a linear number of nodes removed.


Figure \ref{fig:ex} (c) is a scatter plot comparing the size of compression obtained by applying {\FORPI} before {\GFOLU} versus {\GFOLU} before {\FORPI}. Data obtained from the TPTP data set is marked in red; the remaining points are obtained from randomly generated proofs. Points that lie on the diagonal line have the same size after each combination. There are 249 points beneath the line and 326 points above the line. Therefore, as in the propositional case \cite{LURPI}, it is not a priori clear which combination will compress a proof more. Nevertheless, the distinctly greater number of points above the line suggests that it is more often the case that {\FORPI} should be applied after {\GFOLU}. Not only this combination is more likely to maximize the likelihood of compression, but the achieved compression also tends to be larger.

Figure \ref{fig:ex} (d) shows a plot comparing the difference between the cumulative number of resolutions of the first $x$ input proofs and the cumulative number of resolutions in the first $x$ proofs after compression (i.e. the cumulative number of \emph{removed} resolutions). The TPTP data is displayed in the sub-plot; note that the lines for everything except {\FORPI} largely overlap (since the values are almost identical; cf. Table \ref{tab:results}). Observe that it is always better to use both algorithms than to use a single algorithm. The data also shows that using {\FORPI} after {\GFOLU} is normally the preferred order of composition, as it typically results in a greater number of nodes removed than the other combination. An even better approach is to try both combinations and choose the best result (as shown in the `Best' curve).

{\SPASS} required approximately 40 minutes of CPU time (running on a cluster) to generate all the 308 TPTP proofs. The total time to apply both {\FORPI} and {\GFOLU} on all these proofs was just over 8 seconds on a simple laptop computer. The random proofs were generated in 70 minutes, and took approximately 461 seconds (or 7.5 minutes) to compress, both measured on the same computer.
All times include parsing time. These compression algorithms continue to be very fast in the first-order case, and may simplify the proof considerably for a relatively small cost in time.

\section{Conclusions and Future Work}\label{sec:conclusion}

The main contribution of this paper is the lifting of the propositional proof compression algorithm {\RPI} to the first-order case. As indicated in Section \ref{sec:Challenges}, the generalization is challenging, because unification instantiates literals and, consequently, a node may be regularizable even if its resolved literals are not syntactically equal to any safe literal. Therefore, unification must be taken into account when collecting safe literals and marking nodes for deletion.

We first evaluated the algorithm on all 308 real proofs that the \texttt{SPASS} theorem prover (with only standard resolution enabled) was capable of generating when executed on unsatisfiable TPTP problems without equality. Although the compression achieved by the first-order {\FORPI} algorithm was not as good as the compression achieved by the propositional {\RPI} algorithm on real proofs generated by SAT and SMT solvers \cite{LURPI}, this is due to the fact that the 308 proofs were too short (less than 32 resolutions) to contain a significant amount of irregularities. In contrast, the propositional proofs used in the evaluation of the propositional {\RPI} algorithm had thousands (and sometimes hundreds of thousands) of resolutions. 

Our second evaluation used larger, but randomly generated, proofs. The compression achieved by {\FORPI} in a short amount of time on this data set was compatible with our expectations and previous experience in the propositional level. The obtained results indicate that {\FORPI} is a promising compression technique to be reconsidered when first-order theorem provers become capable of producing larger proofs. Although we carefully selected generation probabilites in accordance with frequencies observed in real proofs, it is important to note that randomly generated proofs may still differ from real proofs in shape and may be more or less likely to contain irregularities exploitable by our algorithm. Resolution restrictions and refinements (e.g. ordered resolution 
\cite{hsiang1991proving, OrderedRes}, hyper-resolution \cite{HyperResolution,robinson1965automatic}, unit-resulting resolution \cite{UnitResultingResolution,prover9-mace4}) may result in longer chains of resolutions and, therefore, in proofs with a possibly larger height to length ratio. As the number of irregularities increases with height, such proofs could have a higher number of irregularities in relation to length.

In this paper, for the sake of simplicity, we considered a pure resolution calculus without restrictions, refinements or extensions. However, in practice, theorem provers do use restrictions and extensions. It is conceptually easy to adapt the algorithm described here to many variations of resolution. For instance, restricted forms of resolution (e.g. ordered resolution, hyper-resolution, unit-resulting resolution) can be simply regarded as (chains of) unrestricted resolutions for the purpose of proof compression. The compression process would break the chains and change the structure of the proof, but the compressed proof would still be a correct unrestricted resolution proof, albeit not necessarily satisfying the restrictions that the input proof satisfied. In the case of extensions for equality reasoning using paramodulation-like inferences, it might be necessary to apply the paramodulations to the corresponding safe literals. Alternatively, equality inferences could be replaced by resolutions with instances of equality axioms, and the proof compression algorithm could be applied to the proof resulting from this replacement. Another common extension of resolution is the splitting technique \cite{WeidenbachSplitting}. When splitting is used, each split sub-problem is solved by a separate refutation, and the compression algorithm described here could be applied to each refutation independently.






\begin{footnotesize}
\bibliographystyle{plain}
\bibliography{biblio}
\end{footnotesize}
\end{document}